\documentclass{article}

\usepackage{PRIMEarxiv}

\usepackage[utf8]{inputenc} 
\usepackage[T1]{fontenc}    
\usepackage{hyperref}       
\usepackage{url}            
\usepackage{booktabs}       
\usepackage{nicefrac}       
\usepackage{microtype}      
\usepackage{lipsum}
\usepackage{fancyhdr}       
\usepackage{graphicx}       
\graphicspath{{media/}}     
\usepackage{float}
\usepackage{amsmath,amssymb,amsfonts}
\usepackage{amsthm}
\usepackage{algorithm}
\usepackage{multirow}
\usepackage{bbm}
\usepackage{subcaption}

\title{Reinforcement Learning in Agent-Based Market Simulation: Unveiling Realistic Stylized Facts and Behavior}
\author{Zhiyuan Yao, Zheng Li, Matthew Thomas, Ionut Florescu \\
Stevens Institute of Technology\\
Hoboken, New Jersey, USA\\
\texttt{\{zyao9, zli149, mthomas3, ifloresc\}@stevens.edu}}

\begin{document}

\maketitle

\begin{abstract}

    Investors and regulators can greatly benefit from a realistic market simulator that enables them to anticipate the consequences of their decisions in real markets. However, traditional rule-based market simulators often fall short in accurately capturing the dynamic behavior of market participants, particularly in response to external market impact events or changes in the behavior of other participants. In this study, we explore an agent-based simulation framework employing reinforcement learning (RL) agents. We present the implementation details of these RL agents and demonstrate that the simulated market exhibits realistic stylized facts observed in real-world markets. Furthermore, we investigate the behavior of RL agents when confronted with external market impacts, such as a flash crash. Our findings shed light on the effectiveness and adaptability of RL-based agents within the simulation, offering insights into their response to significant market events.
    
\end{abstract}
\section{Introduction}

Modern financial markets serve as vehicles for setting up-to-date prices, thus shaping economic landscapes worldwide. Thus understanding how markets react to external and internal events is crucial for investors and regulators. Designing a proper financial market simulator can answer ``what if'' questions and help market participants make informed decisions in a fast-paced and volatile market. An extensive body of literature exists dedicated to methods of simulating market behavior \cite{cont2001empirical, cont2010stochastic, rocsu2009dynamic, raberto2001agent, streltchenko2005multi, ardon2021towards, lussange2021modelling}. Among these papers, agent-based market simulators stand out due to their ability to emulate dynamics of the real-world markets. Conventional agent-based systems use rule-based agents, e.g., \cite{wellman2017strategic, farmer2005predictive}. These systems have issues when calibrating to real markets and fail to capture realistic market dynamics. This limitation arises from the rigid, hard-coded nature of rule-based agents, which prevents them from adapting to changing market conditions. In contrast, agents which are capable of learning have the ability to optimize their goals by learning from the environment and the behavior of other agents. This adaptability closely mirrors the behavior of real-world investors, enhancing the realism of the simulation.


Recently, we have seen several successful applications of machine learning techniques in financial problems such as portfolio management \cite{moody1998performance, sun2023trademaster}, credit rating \cite{tsai2010credit, wang2023sparsity}, and order execution \cite{nevmyvaka2006reinforcement, moallemi2022reinforcement}. Reinforcement learning (RL) is an important class of machine learning methods, where agents are capable of learning optimal strategies without knowing the underlying environment dynamics \cite{sutton2018reinforcement}. Recently, several papers have been published that use RL agents to construct a simulation environment for financial markets. Lussange et al. \cite{lussange2021modelling} model a market using hundreds of RL agents, each solving a simplified investment problem. However, the participants of real-world stock markets have different goals and use complex strategies. Ideally, we should let these agents learn to trade in a highly realistic environment where each agent optimizes its own utility. Our proposed simulation framework allows agents to learn complex strategies. In another direction, \cite{ardon2021towards, coletta2022learning} formulate RL-based market makers and liquidity takers and use them to simulate a dealer market. Our work extends this idea to simulate a complete continuous double auction stock market using complex RL-based agents. 


In this study, we propose a simulation framework with only a small group of representative RL agents. We compare this system with one composed of rule-based zero-intelligence agents as well as with real market data. The results obtained using the RL agents' system are comparable with real data. Further, we show that the system is capable of adapting to changing market conditions.  





\section{Important Concepts}
\subsection{Reinforcement Learning Agents}
Mathematically, each RL agent solves a problem associated with a Markov Decision Process (MDP)\footnote{See general formulation in  \cite{sutton2018reinforcement, bertsekas2019reinforcement}.}. A MDP is defined as a tuple $(S, A, R, P, \gamma)$ with several key components:
\begin{itemize}
    \item $S$ is the state space, in our case a set of vectors describing the market limit order book and the agent's account information,
    \item $A$ is the action space which defines the specific orders agents can place. 
    \item $R$ denotes the reward function which specifies the immediate reward for taking an action in response to a particular state,
    \item $P$ denotes the transition probability function, which outputs the probability of transition from one state to another by executing a given action.
    \item $\gamma\in(0, 1)$ is the discount factor; a smaller discount factor lets the agent focus more on recent reward. 
\end{itemize}
When using \textit{model-free} RL methods such as in \cite{mnih2015human, schulman2017proximal}, the dynamics of the system (i.e., the transition probability function $P$) can be unknown. If we denote the policy function of the RL agent as $\pi$, the agent solves the following problem
$$\max_{\pi} \mathbb{E}_{(s_t, a_t)\sim (P,\pi)}\left[\sum_{t=0}^{\infty}\gamma^tR(s_t, a_t)\right].$$

We choose the Proximal Policy Optimization (PPO \cite{schulman2017proximal}) method to optimize our RL agents. 

\subsection{Limit Order Book (LOB) in a Continuous Double Auction (CDA) Market }
Almost all traditional financial exchanges today use a Continuous Double Auction market model. A continuous double auction (CDA) market allows traders to place buy and sell orders continuously at any time \cite{smith2003statistical}. The CDA market maintains two limit order books (LOBs), one for buy orders and one for sell orders. Each order is an instruction placed by a trader who wants to buy or sell an asset at a specific or better price. Since the instruction contains a range of prices for execution this type of order is called a limit order. A market order is an instruction to buy or sell an asset immediately at the current market price. Generally, limit orders stay in the LOB until they are matched with an incoming market order.   

\section{System and Agents}
\subsection{System Description}
The system contains a machine engine that organizes LOBs and settles trades, as well as a brokerage center that keeps track of each agent's account, including the agent's buying power and assets. All agents place market and limit orders to the matching engine through their brokerage accounts. The matching engine runs a CDA market model. The engine updates the latest LOB information and streams its state to each trading agent in real time. 

The agents in this system are of two types: liquidity-taking (LT) agents and market-making (MM) agents. Each instance of these agents is formulated as an RL agent, each with its own parameters and reward function. Each agent observes the system independently, selects actions, receives feedback, and optimizes its own strategy. Each agent learns to adapt its strategy through actions (orders submitted) and feedback received (reward). The formulation of rewards is different for each agent, we provide details in the next section.

We \textit{highlight two aspects of our work} which we think helped improve the realism of the simulation compared to prior work. First, all agents run in their own respective threads, thus all threads run in parallel and are not waiting for any other thread once they are launched. Second, all agents are heterogeneous. Even though some agents belong to the same category, they use different sets of hyperparameters, and this results in significantly different behavior for each agent.

\subsection{Agents}

Our formulation for the MM and LT agents is based on the formulation provided in  \cite{ardon2021towards}. The scope of the original paper was to simulate a dealer market rather than a CDA stock market. In the original paper the quotes are provided by multiple RL agents called liquidity providers (dealers) and by an Electronic Communication Network (ECN) agent. In practical implementation, the ECN is a replay of market quotes from a particular day. The other class of RL agents are liquidity takers which have an investment goal defined by their own reward function. The focus of the paper is to understand the interactions among the two types of agents. Specifically, they want to understand how the agents cooperate in the presence of an external public order flow.  

In our simulation, the ECN is gone. This allows us to diverge from the \textbf{one reality problem} inherent in financial simulators. When the simulator replays market activity there is a single predetermined price path. Asset prices may diverge slightly but they will all come back to the real history. In the original work, all agents use a shared policy which they learn collaboratively. We believe this approach works and it is relevant in their case, precisely due to the fixed flow ECN. In our case, the shared policy creates behavior that is quite correlated despite the fact that each agent has its own set of hyper-parameters as well as reward structure. This is why in our implementation each agent has its own policy function. 

Since our system does not have a stream of orders guiding the evolution of asset prices, we have to establish the realism of the simulated market activity. Only in a realistic market can we try to understand RL agents' behavior. We present the specific formulations for both MM and LT agents in the following subsection.

\subsubsection{Market-Making (MM) Agent}
~\\
\noindent\textbf{Observation Space.} Each MM agent observes: mid-prices for the past 5 time steps, prices and quantities of top 5 levels of the LOB, liquidity provision percentage, current inventory, and its buying power. 
The liquidity provision percentage for MM agent $i$ is measured by
$$P^{i} = \frac{\sum_{l=1}^d m_l^{(i)}}{\sum_{i=1}^N\sum_{l=1}^5 m_l^{(i)}}$$
where $m_l^{(i)}$ is the amount of shares placed in the LOB by agent $i$ at price level $l$, $d$ and $N$ are the total number of price levels and the total number of MM agents, respectively. 
Hyper-parameters in the reward function are also appended to the observation vector.

\begin{figure}
    \centering
    \includegraphics[width=0.4\textwidth]{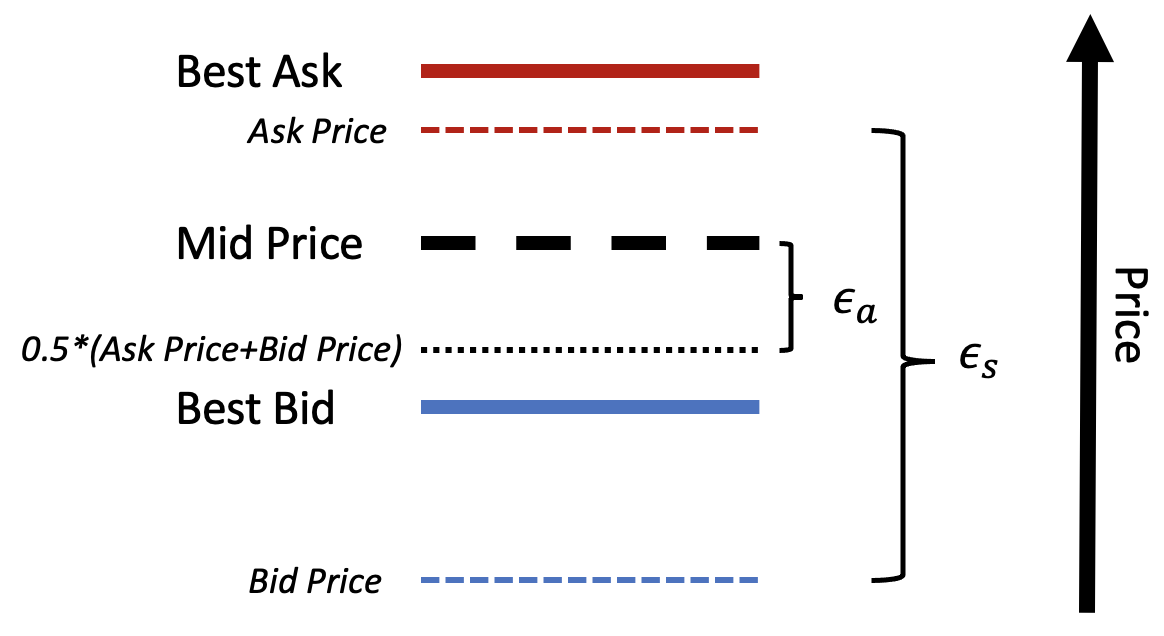}
    \caption{A demonstration of the formulation of market makers' action. }
    \vspace{-1em}
    \label{fig:MM_action_demo}
\end{figure}

\noindent\textbf{Action Space.}
At each time step, the MM agent places two limit orders on both sides of the LOB. The policy function of the MM agent generates three values that collectively determine both the price and size of these limit orders:
\begin{itemize}
    \item A percentage of the buying power, which is used to determine the size of the limit orders,
    \item The symmetric price tweak $\epsilon_s$,
    \item The asymmetric price tweak $\epsilon_a$.
\end{itemize}
The symmetric and asymmetric price tweaks determine the prices of both bid and ask orders (See illustration in Figure \ref{fig:MM_action_demo}). 
The price of the bid order and ask order follow the formula 
\begin{align}
    p_{\text{ask}} =& p_{\text{mid}}+s\cdot\left((1+\epsilon_s)/2+\epsilon_a\right)\label{eq:asymtweak}\\
    p_{\text{bid}} =& p_{\text{mid}}-s\cdot\left((1+\epsilon_s)/2-\epsilon_a\right)\nonumber
\end{align}
where $s$ is the spread between the best bid and ask prices. The symmetric price tweak controls the distance between the ask price and the bid price. The asymmetric price tweak controls the average of bid and ask prices moving up/down (see Figure \ref{fig:MM_action_demo}). 
We follow the original formulations in \cite{ardon2021towards}.

\noindent\textbf{Reward}
The reward function for MM agents follows
\begin{align}
    r^{(\text{MM}_i)}_t &= \omega\left(\alpha(\Delta \text{PnL}_t-\gamma*|\Delta \text{PnL}_{\text{inventory},t}|)\right)\nonumber\\
    &\quad -(1-\omega)\left|P_t^i-P_{t}^{*, i}\right| \label{eq:mmreward}
\end{align}
where $\omega$ is the weight for the PnL component , $\alpha$ the PnL component normalizer, the inventory risk sensitivity $\gamma$, and the target liquidity provision percentage $P_{t}^{*, i}$ are all hyper-parameters of the MM agent $i$.

The reward structure in \eqref{eq:mmreward} includes both the PnL as well as the amount of liquidity provided by the agents. The first term in the reward function provides an incentive to increase the PnL while imposing penalties for PnL fluctuations resulting from outstanding inventory and price oscillations. The second term in the reward function aims to minimize the discrepancy between the actual liquidity provision percentage and the target liquidity provision percentage.

\subsubsection{Liquidity Taking (LT) Agent}
~\\
\textbf{Observation Space} LT agents have the same observation space as the MM agents without the liquidity provision percentage.

\noindent\textbf{Action Space}
At each step, an LT agent can choose to place a bid or an ask order, or do nothing and skip this step. Once the order type is chosen, the agent will send a market order with a fixed order size. 

\noindent\textbf{Reward}
The reward function for LT agents follows
\begin{align}
    r^{(\text{LT}_i)}_t &= \omega(\alpha(\Delta \text{PnL}_t-\gamma|\Delta \text{PnL}_{\text{inventory}, t}|)) \label{eq:ltreward}\\
    -&\frac{(1-\omega)}{2}(\Delta|f_{\text{buy}}^{*, i}-n_{\text{buy}}^{(t-\tau,t)}/\tau|+\Delta|f_{\text{sell}}^{*, i}-n_{\text{sell}}^{(t-\tau, t)}/\tau|)\nonumber
\end{align}
where $\omega$ is the weight for the PnL component, $\alpha$ is the PnL component normalizer, $\gamma$ adjusts the inventory risk sensitivity, and the target buy and sell fractions ($f_{\text{buy}}^{*, i}$, $f_{\text{sell}}^{*, i}$) are hyper-parameters of the LT agent $i$.

Similar to the reward function of MM agents, this reward structure also contains two parts: a PnL component and an order frequency component; the weight $\omega$ adjusts the respective contribution of the two components. The first term is the same as the term in MM's reward function. The second term of the reward function calculates the difference between the target order frequency and the actual order frequency. The buy order frequency is $n_{\text{buy}}^{(t-\tau,t)}/\tau$, where $n_{\text{buy}}^{(t-\tau,t)}$ is the total number of orders bought from time $t-\tau$ to time $t$, and $\tau$ is a hyper-parameter indicating the length of the time window. Similar to the sell order frequency $n_{\text{sell}}^{(t-\tau,t)}/\tau$.

\subsection{Simulation Details}
We initialize the MM and the LT agents with a random amount of buying power and assets. All hyper-parameters of MM and LT agents are randomly sampled. 
When a simulation starts, all agents are launched using their individual threads. 

At the beginning of each time step, all MM and LT agents observe the system and send market or limit orders to the system. 
Each agent collects experience and stores it in an individual dataset. Training of the agent takes place independently after a certain amount of data has been collected. We train the agents using Proximal Policy Optimization (PPO) \cite{schulman2017proximal}.

The simulation is implemented within the SHIFT system \cite{alves2020shift}, a real-time high-frequency trading platform. The SHIFT system allows clients to connect through the FIX protocol across the Internet, which realistically replicates the connectivity in modern exchanges. Additionally, random latency due to network communication introduces an element of true randomness to the simulation, enhancing its realism.  





%


 \section{Experiment Design} 

In order to provide realistic answers to questions about RL behavior, it is crucial to ensure that the simulation is similar to a real market. 
We list a set of properties inherent in real-world financial markets. We then analyze the results of the simulations to see whether they exhibit these properties. 

In general, RL agents assume unchanged dynamics of the system. However, in real-world scenarios when interacting with other traders and in a dynamic market, the system dynamics evolve. Assuming that the system is realistic,  we investigate whether training RL agents during the simulation improves their behavior and enhances their adaptability to changing market conditions.

The following two sub-sections discuss in detail the design of the experiment to answer these two questions:
\begin{enumerate}
    \item Can RL agents simulate a realistic market?
    \item Should agent training persist during simulation?
\end{enumerate}

\subsection{Desired Market Properties}


A realistic simulation should look and behave like a real-world market. To analyze and quantify a realistic behavior we examine two aspects: statistical characteristics and market responsiveness. Statistical characteristics include distributional properties such as the price/return distribution, as well as time dynamics such as auto-correlations of the return series, volatility clustering, and more. Market responsiveness measures changes in several variables in response to a large sell order hitting the market. 



\subsubsection{Statistical Characteristics}
Prior research using empirical financial time series data found some statistical properties that are shared by a wide range of assets and across various time periods. For instance, stock returns have a distribution that has a sharper peak and fatter tail, compared to a normal distribution. These statistical characteristics are known as \textit{stylized facts} \cite{cont2001empirical}. We examine whether the results of our simulation show these stylized facts previously evidenced in literature \cite{cont2001empirical, ratliff2023revisiting}. We focus on the following well-known single-asset stylized facts:
\begin{enumerate}
    \item Heavy tails and kurtosis decay with increasing $\Delta t$,
    \item Absence of auto-correlations,
    \item Slow decay of auto-correlation in absolute returns,
    \item Volatility clustering.
\end{enumerate}

In addition to these variables which are the result of the entire market activity, we examine individual market makers' performance records and the evolution of their inventory and PnL component. MM agents should have a stable inventory evolution and the absolute value of the inventory should be close to 0. 
MM agents should make most of their profit from the spread rather than the inventory PnL. 

\subsubsection{Market Responsiveness}
A simulated market within a closed environment where trading is driven by traders without any outside information should be stable. However, we are interested in how the simulation will behave when we additional external impact is introduced. We want to see if changes in behavior align with documented literature on real-world cases. In a real-world market, a large liquidation typically causes price impact. Almgren and Chriss \cite{almgren2001optimal} assert that price impact manifests in two forms: temporary impact and permanent impact. Therefore, following a substantial sell order, the price is observed to undergo a transient decline, followed by a recovery to a level that remains below the pre-sale price. 
Additionally, a large directional order flow carries information, such as the potential for substantial liquidation. Market makers can capture the information from their order flows and adjust their strategies accordingly. In practical scenarios, when more orders come from one side than the other side, market makers tend to widen the spread and shift the center of price spread towards the direction which has fewer orders \cite{glosten1985bid, copeland1983information, putnam2018describing}. 
We examine the behavior of the MM agents to assess whether similar patterns are observed. To do this, we design two experiments with different types of informed traders. The first experiment introduces a flash-sale agent who places large sell orders to the simulated market. The second experiment involves dynamically altering the buy/sell preferences of LT agents throughout the simulation, prompting them to trade in the same direction within short time periods, thereby influencing price movements.



\subsection{Continual Learning}

We introduce three groups of agents in the simulation. 


\begin{itemize}
    \item Group A - Continual Training Group. The agents are pre-trained for 10 hours (36,000 steps), and training continues throughout the time of the simulation (for another 10 hours or 36,000 steps). 
    \item Group B - Testing Group. The agents in this group are pre-trained for 10 hours and are used in the simulation \textit{without} continuing training.
    \item Group C - Untrained Group. The third group serves as a control to understand the performance improvement obtained from training. The agents in this group load the random initialized parameters and run simulations without training.
\end{itemize}

\begin{figure}[h]
    \centering
    \includegraphics[width=0.60\textwidth]{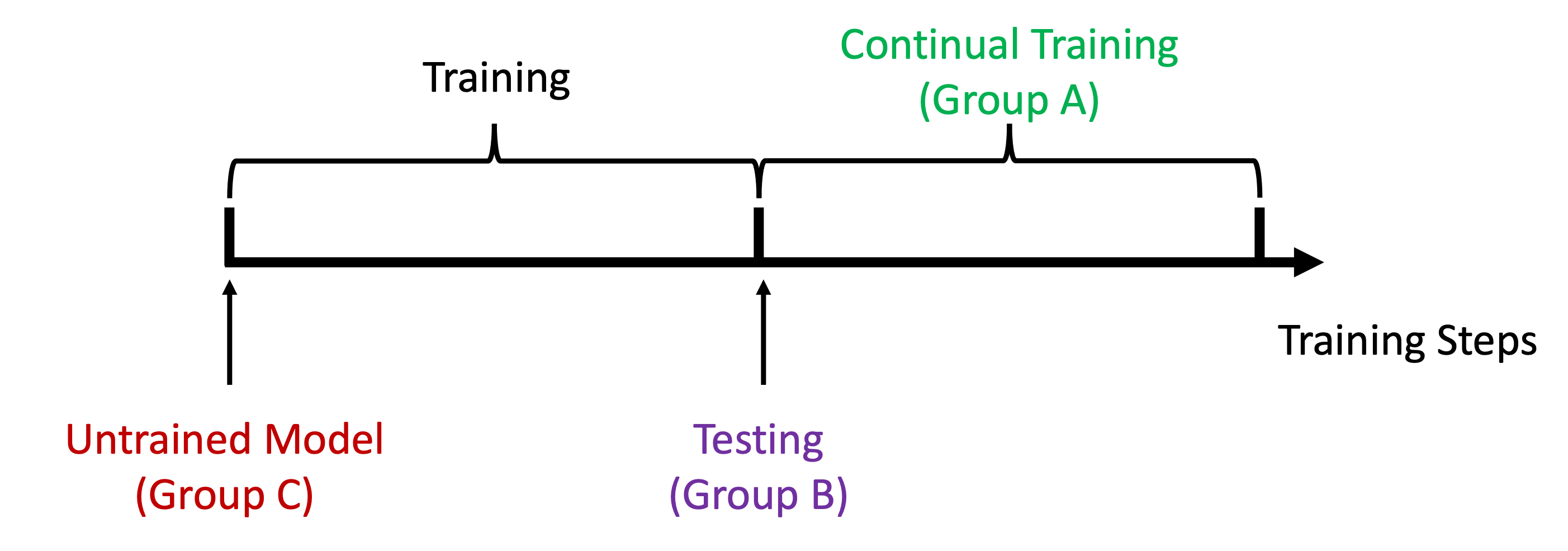}
    \caption{Illustration of Groups for Comparison}
    \label{fig:groups_ill}
\end{figure}

For each random seed, we generate the parameters of the neural networks for the Group C agents directly. Each agent in Group C is trained for 10 hours and their parameters become the parameters used for each agent in Group B. The same parameters are used to initialize the agents in Group A. We are describing the process in detail as this is similar to a matched pairs testing design to minimize randomness for comparison purposes. This is important because we only repeat this process for 10 random seeds, that is 10 simulations. Each simulation takes 20 hours when running all of them in parallel and there are a lot of computational resources required for this study. This process is illustrated in Figure \ref{fig:groups_ill}.

To compare the results produced by the RL agents we introduce an additional simulation model using 100 Zero-Intelligence (ZI) agents. This system is using the agent design in Farmer et al.'s work \cite{farmer2005predictive} and \cite{alves2020shift}. We analyze and compare stylized facts obtained using the RL agents system, the ZI agents system, and real data. Additionally, we investigate the evolution of Market-Maker (MM) agents' inventory and PnL components across different groups. To assess responsiveness, we introduce a sequence of flash sale events, and we examine the price impact during the flash sale period. We also examine the MM's change in behavior due to the flash sale, and we evaluate the adaptability of continual learning agents by comparing policies before and after training with the flash sale prices.

\section{Experiment Results}

\subsection{Statistical Characteristics of Observed Asset Prices}
We use Google, Apple, and Amazon’s tick level limit order book data on 2012-06-21 from LOBSTER \cite{lobster}. We are presenting graphs for Group A only due to page limitations. The continual learning Group A is the most interesting one to analyze. 

\begin{figure}[ht]
    \centering
    \begin{subfigure}[b]{0.7\textwidth}
        \includegraphics[width=\textwidth]{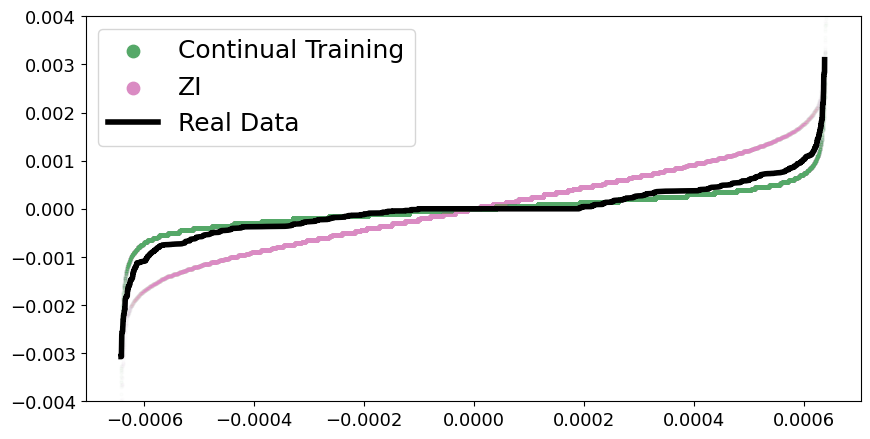}
        \caption{Quantile-Quantile plot for (10-second) return distributions simulated by RL agents and ZI agents, compared against the distribution of the real data.}
        \label{fig:rl_zi_qqplot_with_goog}
    \end{subfigure}

    \begin{subfigure}[b]{0.35\textwidth}
        \includegraphics[width=\textwidth]{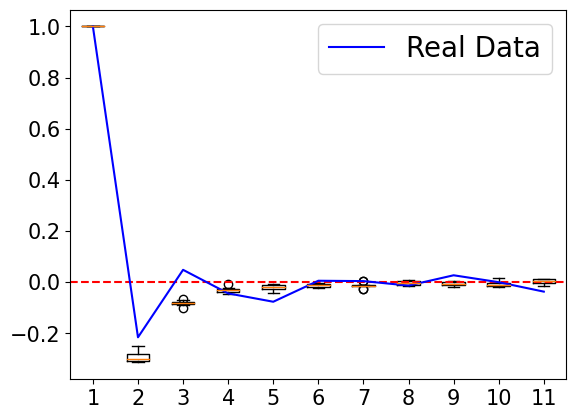}
        \caption{RL-agent Simulation}
        \label{fig:train_autocorr_with_goog}
    \end{subfigure}
    \begin{subfigure}[b]{0.35\textwidth}
        \includegraphics[width=\textwidth]{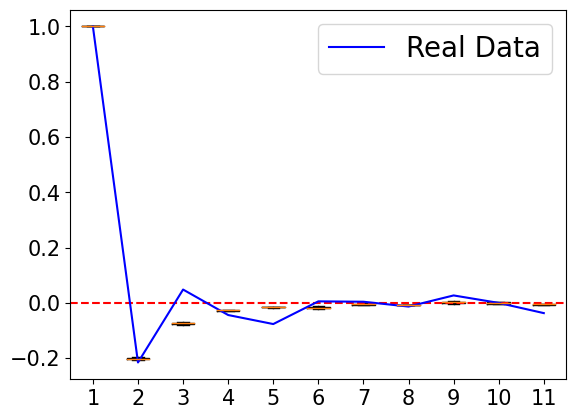}
        \caption{ZI-agent Simulation}
        \label{fig:ZI_autocorr_with_goog}
    \end{subfigure}

    \caption{Returns' QQ plot \& auto-correlation comparison between RL and ZI.}
    \label{fig:rl_zi_autocorr_analysis}
\end{figure}

\textbf{Heavy tails and kurtosis decay.} We measure leptokurtic behavior using excess kurtosis which should be significantly larger than 0. Decay means that kurtosis decreases as the sampling frequency of returns decreases (seconds, minutes, hours, etc.). Figure \ref{fig:rl_zi_autocorr_analysis} shows the Quantile-Quantile plots of simulations for the continual training RL agents (Group A) and ZI agents respectively, compared with real data. 
The real data, as well as data obtained using RL and ZI agents, all exhibit strong fat tail return and price distributions. The results from the ZI agents simulation show a milder tail. Results from RL agents closely align with the real data. The average kurtosis from the 10 simulations is presented in Table \ref{tab:kurtosis decay}. The table shows the kurtosis decays as data is sampled less frequently. 

\textbf{Absence of auto-correlations.} \cite{cont2001empirical} describes this property as small insignificant values for auto-correlation of the return time series unless within a very short time interval. Figure \ref{fig:rl_zi_autocorr_analysis} shows a strong negative correlation in the first lag and that the auto-correlation function decays to 0 for increasing lags. The tighter boxplots in \ref{fig:rl_zi_autocorr_analysis} imply the ZI agents generate a very consistent market, while the real data and RL agents exhibit more variability in this analysis. The author of \cite{cont2001empirical} cites the bounce of market orders between bid and ask prices as the reason why the first lag shows stronger negative auto-correlation.

\begin{figure}[ht]
  \centering
  \begin{subfigure}[b]{0.35\textwidth}
    \centering
    \includegraphics[width=\textwidth]{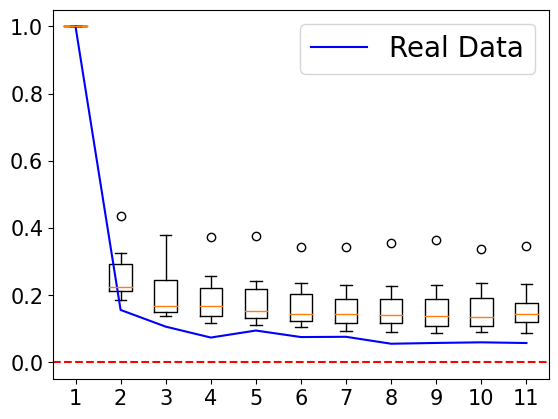}
    \caption{RL-agent Simulation}
    \label{fig:absolue_autocorr_train}
  \end{subfigure}
  \begin{subfigure}[b]{0.35\textwidth}
    \centering
    \includegraphics[width=\textwidth]{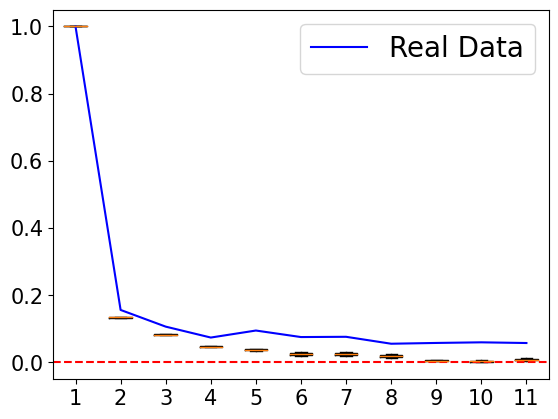}
    \caption{ZI-agent Simulation}
    \label{fig:absolue_autocorr_ZI}
  \end{subfigure}
  \caption{Auto Correlation of Absolute Returns}
  \label{fig:Absolute return auto correlation anaylysis}
\end{figure}

\textbf{Slow decay of auto-correlation for absolute returns.} The auto-correlation function of (1-second) absolute returns decays slowly as a function of the time lags, this implies a long-range time dependency of return magnitude. Figure \ref{fig:Absolute return auto correlation anaylysis} shows auto-correlation of absolute returns from the simulations using RL agents (Figure \ref{fig:absolue_autocorr_train}) and the ones using ZI agents (Figure \ref{fig:absolue_autocorr_ZI}). The blue lines in both plots indicate the autocorrelation of the absolute returns of the real data. It can be seen that the auto-correlations from all three methods are decreasing slowly with increasing lags. Figure \ref{fig:absolue_autocorr_ZI} shows again tighter boxes indicating the simulation using ZI agents is more consistent. Comparing both plots, we can see that RL agents simulate a market in which the return magnitude has a larger autocorrelation than the market simulated by ZI agents.  

\begin{figure}[ht]
    \centering
    \begin{subfigure}[b]{0.35\textwidth}
        \includegraphics[width=\textwidth]{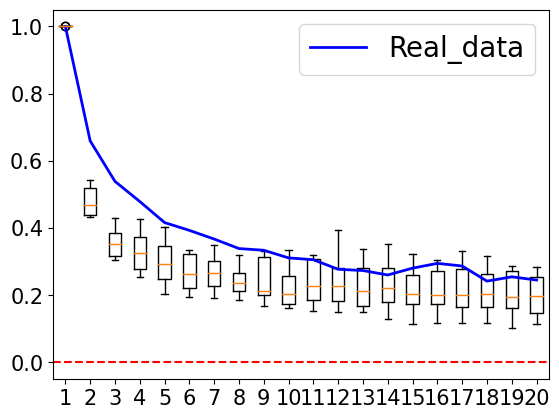}
        \caption{RL-agent Simulation}
        \label{fig:vol_cluster_train}
    \end{subfigure}
    \begin{subfigure}[b]{0.35\textwidth}
        \includegraphics[width=\textwidth]{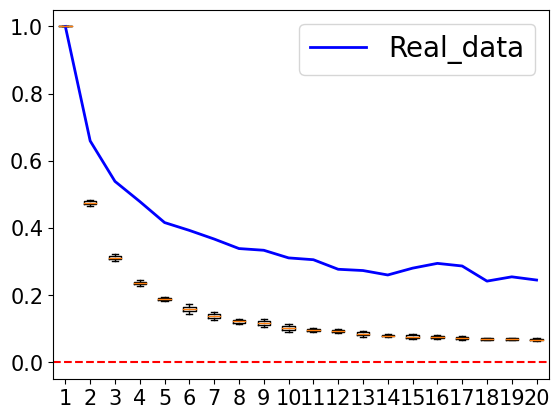}
        \caption{ZI-agent Simulation}
        \label{fig:vol_cluster_ZI}
    \end{subfigure}
    
    \caption{Volatility Clustering Analysis}
    \label{fig:Volatility Clustering Analysis}
\end{figure}

\textbf{Volatility clustering.} Cont et al. \cite{cont2001empirical} use autocorrelation of \textit{squared} return time series to measure the volatility clustering effect. Figure \ref{fig:Volatility Clustering Analysis} shows decreasing auto-correlations in both types of simulation. This indicates that they all exhibit volatility clustering effects. Similar to the auto-correlations of absolute returns, the auto-correlation in the ZI-agent simulated market shown in Figure \ref{fig:vol_cluster_ZI} is significantly lower than that of the real data, indicating a less pronounced clustering effect in the ZI-agent simulated market than the RL-agent simulated market.


\begin{table}[ht]
  \centering
  \caption{Kurtosis As Return's $\Delta t$  increases}
    \begin{tabular}{|l|c|c|c|c|}
    \hline
    \multirow{2}{*}{\textbf{Groups}} & \multicolumn{4}{c|}{\textbf{$\Delta t$}} \\ \cline{2-5}  
     & \textbf{1s} & \textbf{30s} & \textbf{60s} & \textbf{120s} \\
    \hline
    \textbf{Real Data} & 92.26 & 5.11 & 3.51 & 2.41 \\
    \hline
    \textbf{ZI} & 3.15 & 1.13 & 1.23 & 0.84 \\
    \hline
    \textbf{Cont. Training} & 10.24 & 6.68 & 5.29 & 3.76 \\
    \hline
    \textbf{Testing} & 142.65 & 117.87 & 98.11 & 54.01 \\
    \hline
    \textbf{Untrained} & 74.25 & 38.12 & 26.65 & 16.86 \\
    \hline
  \end{tabular}
  \label{tab:kurtosis decay}
\end{table}

\begin{figure}[h]
  \centering
  \includegraphics[width=0.7\textwidth]{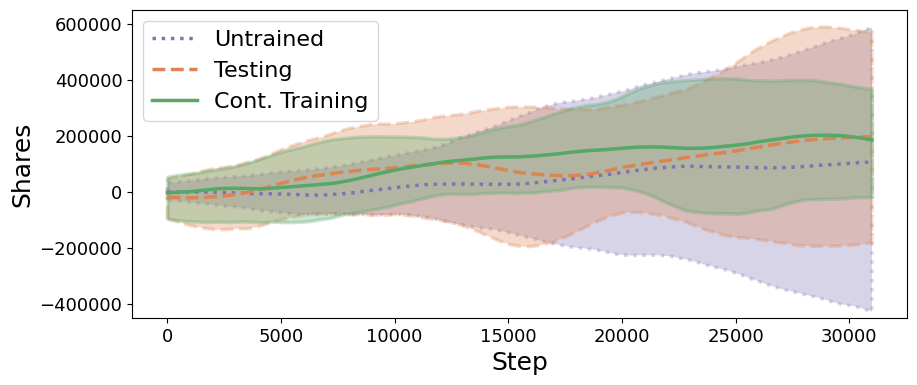}
  \caption{Market Making Agents' Inventory Changes. The plot shows the mean evolution of inventories of the three groups. The band shows 20-80 percentile of inventory from all MM agents. 
  }
  \vspace{-1em}
  \label{fig:MM inventory (20-80 percentile)}
\end{figure}

To assess realistic behavior, the MM agents should control their inventory and rely on the Profit and Loss (PnL) from spread (providing liquidity) rather than from inventory (long-term investments). Figure \ref{fig:MM inventory (20-80 percentile)} shows the average inventory evolution of all MM agents in the three groups over time. The continual training group has the smallest variation in the inventory, while the untrained group has the largest variation. We further separate the MM's PnL into profit from spread and profit from holding inventory. The MMs in the continual training group profit on average \$1.26 million from spread and lose \$0.22 million from inventory. The results for MM agents from other groups are similar. 

\subsection{Market and Agent Responsiveness to External Events}
In the first experiment, we introduce a rule-based agent which performs a sequence of ``flash sales''. During a flash sale event, this agent continuously sends sell orders (300 lots each) every second for a duration of 5 seconds. Then, the agent remains inactive for 400 seconds to allow the market to recover. The goal is to study how the market moves during and after the flash sale events and how the agents' behavior changes after training with these events. 

\begin{figure}[ht]
  \centering
  \begin{subfigure}[b]{0.7\textwidth}
    \centering
    \includegraphics[width=\textwidth]{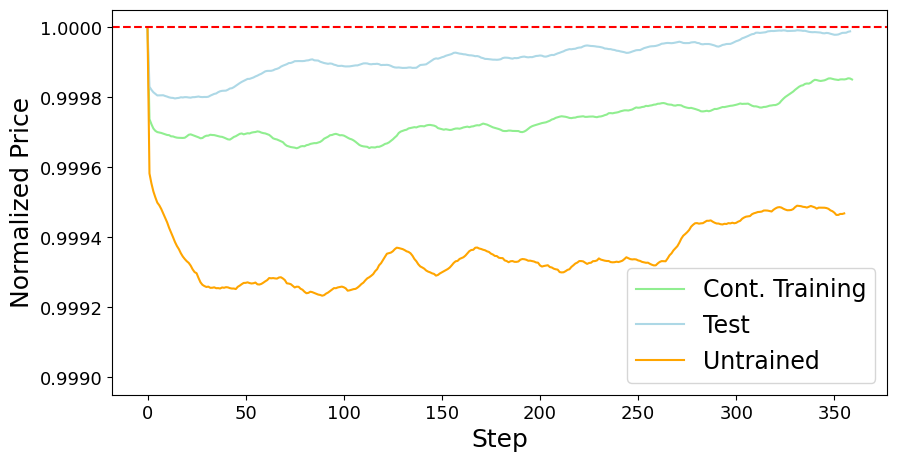}
  \end{subfigure}
  \caption{Price Impact Caused by Flash Sale Orders. The plot shows the price movement over the time steps. The y-axis shows stock prices normalized by the price before the flash sale events.}
  \vspace{-0.5em}
  \label{fig:Flash crashes: Price Impacts}
\end{figure}

Figure \ref{fig:Flash crashes: Price Impacts} depicts the price movement after the start of flash sale events. To create these plots we use the average price of multiple simulations normalized by the price at which flash sale started. The untrained group shows a significant and permanent price drop in contrast to the testing and continual training groups. In the testing group, prices recover to pre-flash-sale levels. However, the continual training group shows more realistic trajectories, with prices dropping and recovering to levels lower than the pre-sale prices. This price pattern aligns with descriptions of permanent and temporary price impact in \cite{almgren2001optimal}.

\begin{figure}[h]
  \centering
  \begin{subfigure}[b]{0.24\textwidth}
    \includegraphics[width=\textwidth]{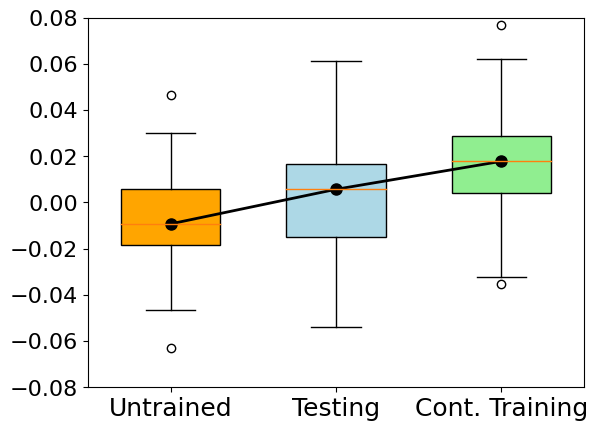}
    \caption{\footnotesize Imbalanced: Action Sym.}
    \label{fig:ib_sym}
  \end{subfigure}
  \begin{subfigure}[b]{0.24\textwidth}
    \includegraphics[width=\textwidth]{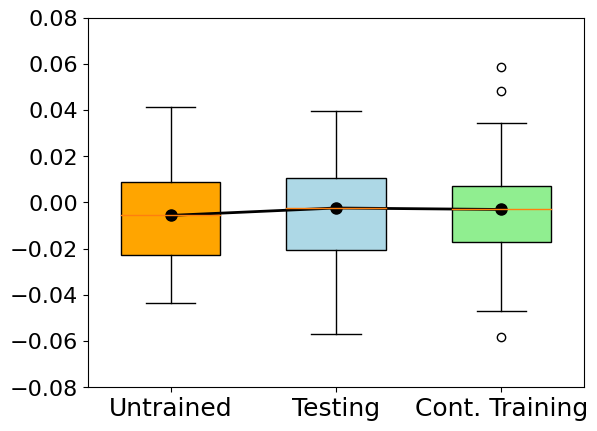}
    \caption{\footnotesize Balanced: Action Sym.}
    \label{fig:b_sym}
  \end{subfigure}
  \begin{subfigure}[b]{0.24\textwidth}
    \includegraphics[width=\textwidth]{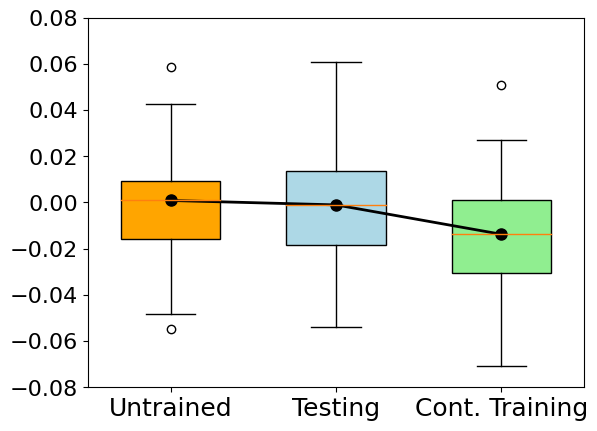}
    \caption{\footnotesize Imbalanced: Action Asym.}
    \label{fig:ib_asym}
  \end{subfigure}
  \begin{subfigure}[b]{0.24\textwidth}
    \includegraphics[width=\textwidth]{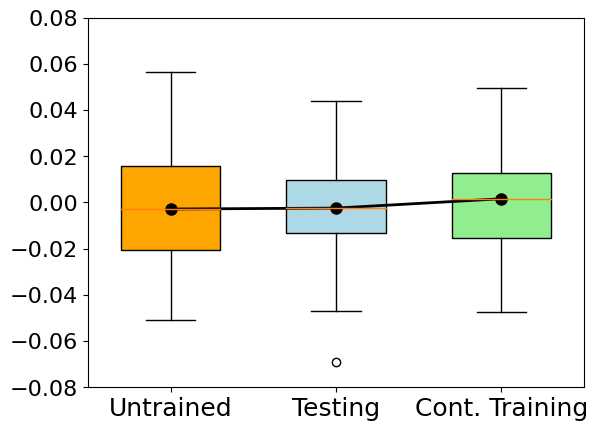}
    \caption{\footnotesize Balanced: Action Asym.}
    \label{fig:b_asym}
  \end{subfigure}
  \caption{MM's Actions Comparison in Flash Sale Events}
  \label{fig:MM's actions comparison}
\end{figure}

We are also interested in how RL agents' policy functions are affected when trained with flash sale events. Recall the action symmetric tweak and asymmetric tweak parameters control the prices of the limit orders sent by MM's (see \eqref{eq:asymtweak} and illustration in Figure \ref{fig:MM_action_demo}). Intuitively, a larger symmetric tweak makes the agent place orders away from the best bid and ask (more conservatively). A positive asymmetric tweak indicates that the MM expects future higher prices than the current one, while a negative value indicates the MM expects the market to go down. When a flash sale event happens, market sell orders take liquidity from the bid side of the LOB and this results in an imbalanced LOB. In contrast, a balanced LOB means the volumes on the buy and sell sides are comparable. To determine the MM agents' reaction we collect the states with balanced and imbalanced LOBs, and feed these states to the different agents' policy functions. Figure \ref{fig:MM's actions comparison} shows the action symmetric tweak values in the top row and the asymmetric tweak in the bottom row. We can see that when the order book is balanced the three groups output a similar distribution of actions (Figure \ref{fig:b_sym} and \ref{fig:b_asym}). However, when fed imbalanced LOBs (Figures \ref{fig:ib_sym} and \ref{fig:ib_asym}), the agents in the continual training group output higher symmetric price tweaks and lower asymmetric price tweaks, compared with the agents in the two other groups. Thus when a flash sale event is detected, the MM agents in the continual learning group tend to place orders deeper into the LOB and move the expected price equilibrium downward. This behavior aligns with the findings in literature \cite{glosten1985bid, copeland1983information, putnam2018describing}. These results show that the agents can adapt to different market conditions through continual learning.

In the second experiment, instead of having information coming from ``flash sales'' we model information flow through informed traders. The informed traders are modeled using LT agents with dynamically changing target buy/sell fractions throughout the simulation. Refer to the reward function for LT agents in \eqref{eq:ltreward}. In this equation, we use two hyperparameters to control the target buy/sell fraction. For example, when agent $i$ has $f^{,i}_{\text{buy}}>f^{,i}_{\text{sell}}$, the agent tends to place more buy orders than sell orders. If all the agents trade in the same direction, the LT agents create a strong imbalanced order flow and drive the price towards a direction. 

\begin{figure}[h]
    \centering
    \includegraphics[width=0.7\textwidth]{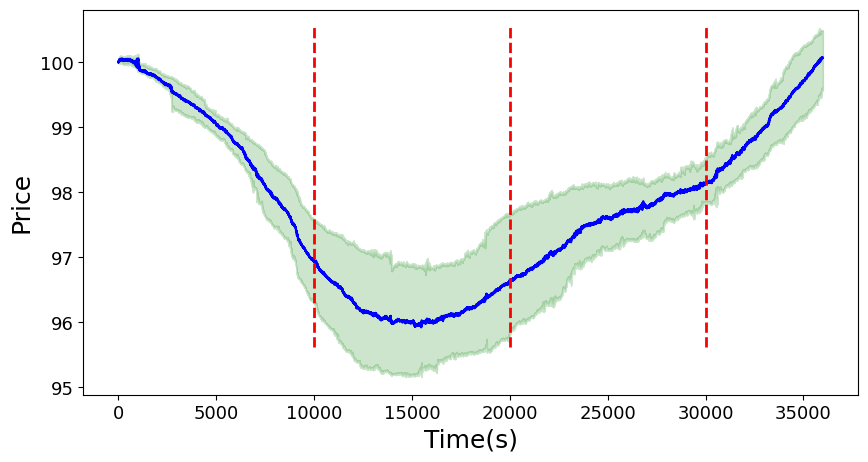}
    \caption{Price Movement with Changing Buy/Sell Preference of LT agents. Results are aggregated from 
    several generated paths. }
    \vspace{-1em}
    \label{fig:Market signals: Price movement}
\end{figure}

\begin{figure}[ht]
    \centering
    \begin{subfigure}[b]{0.24\textwidth}
        \includegraphics[width=0.9\textwidth]{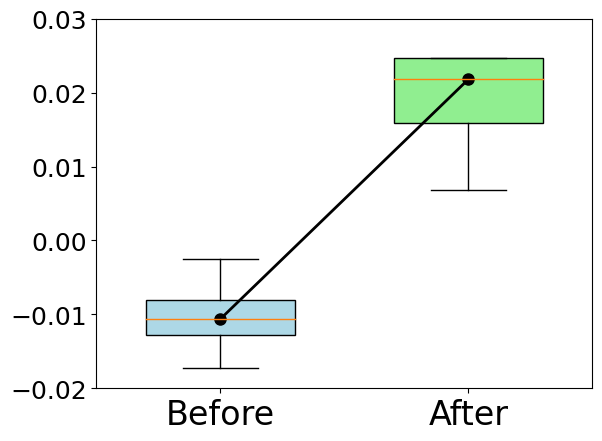}
        \caption{\footnotesize Price Down: Action Sym.}
        \label{fig:all_lt_down_act_sym}
    \end{subfigure}
    \begin{subfigure}[b]{0.24\textwidth}
        \includegraphics[width=0.9\textwidth]{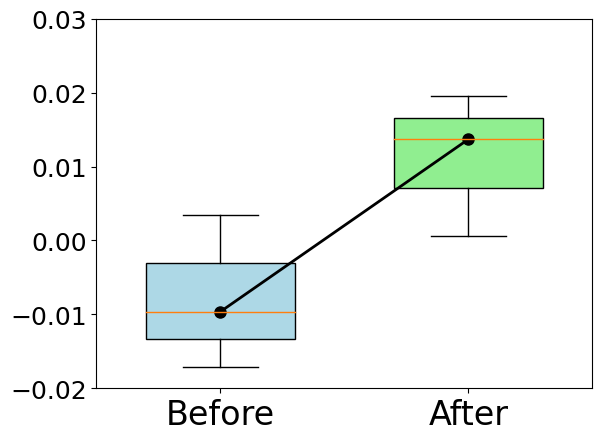}
        \caption{\footnotesize Price Up: Action Sym.}
        \label{fig:all_lt_up_act_sym}
    \end{subfigure}
    \begin{subfigure}[b]{0.24\textwidth}
        \includegraphics[width=0.9\textwidth]{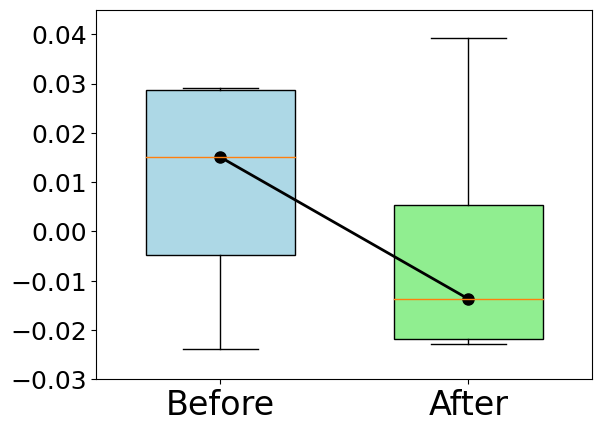}
        \caption{\footnotesize Price Down: Action Asym.}
        \label{fig:all_lt_down_act_asym}
    \end{subfigure}
    \begin{subfigure}[b]{0.24\textwidth}
        \includegraphics[width=0.9\textwidth]{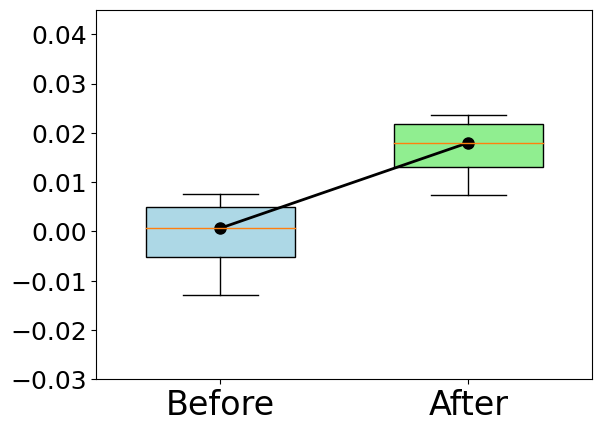}
        \caption{\footnotesize Price Up: Action Asym.}
        \label{fig:all_lt_up_act_asym}
    \end{subfigure}
    \caption{MM's Actions Comparison in Simulation with Informed LT Agents.}
    \vspace{-2em}
    \label{fig:Type II inform traders: MM's Actions Comparison}
\end{figure}

In this experiment, we only use continual learning Group A MM agents, as they need to adapt to changing market conditions. The LT agents are also from Group A but their reward function is changed through the evolution of the target buy/sell parameters. Specifically, Figure \ref{fig:Market signals: Price movement} shows the price process resulting from the activity of these informed LT agents. The four phases separated by dashed red lines are: Sell (0.3/0.4), Buy (0.4/0.35), Balanced buy and sell (0.4/0.4), and last Buy (0.4/0.3), the numbers in parenthesis indicate (buy fraction/sell fraction). 
We expect to see the price movement aligning with the target buy/sell parameters. Additionally, similar to the previous study we collect states in the first and the last phases (i.e., steps 0-10,000 when the price goes down and steps 30,000-36,000 when the price goes up). We feed these states to the MM's policy function at the beginning of the day (Before) and at the end of the trading day (After). Figure \ref{fig:Type II inform traders: MM's Actions Comparison} shows the distributions of the outputs in action symmetric and asymmetric tweak. In both scenarios when the price is going down and when the price is going up, the action symmetric tweak (top row) gets larger, thus the MM agent becomes more conservative and tends to enlarge the spread. This aligns with the findings in the previous experiment. We can also see that the distribution of price asymmetric tweak becomes negative when prices are going down and becomes positive when prices are going up. This means the learned agents change their expectations of future prices along with the observed market direction. In contrast, the asymmetric tweak average is close to zero for agents that are not continuously trained. This helps explain why the price tends to come back to pre-flash-sale values for the agents in Group B in Figure \ref{fig:Flash crashes: Price Impacts}.

\section{Conclusion}
In this work, we modify the formulation of RL agents in \cite{ardon2021towards} and implement a highly realistic simulation platform. We compare the simulation results against a real data set and a market simulated using zero intelligence (ZI) traders. The results obtained using the simulation platform show realistic market characteristics and responsiveness to external factors. We find that continual learning RL agents produce the most realistic market simulation, and are capable of adapting to changing market conditions. 

Calibration of an agent-based system is still a challenging problem. Vadori et al. \cite{vadori2022towards} and Lussange et al. \cite{lussange2021modelling} show two-step procedures to calibrate the RL-based multi-agent system. However, applying these algorithms to our system is very challenging as our system is non-stationary and runs in real time. We plan to address this and other issues in future work.


\newpage
\bibliographystyle{IEEEtran}
\bibliography{paper}

\newpage
\section{Appendix}

\setcounter{figure}{0}
\setcounter{table}{0}

\subsection{Additional Simulation Results}
\begin{figure}[!htb]
    \centering
    \includegraphics[width=0.7\textwidth]{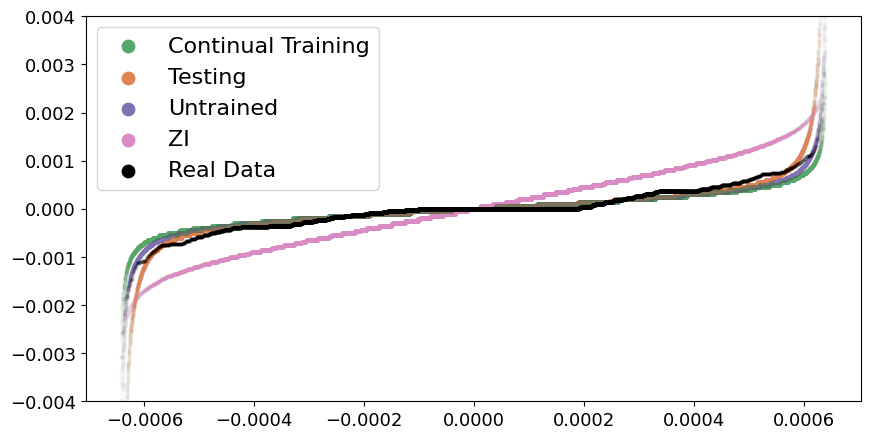}
    \captionsetup{font=large, justification=centering}
    \caption{Quantile-Quantile plot for (10-second) return distributions simulated \\by all agent groups, compared against the distribution of the real data.}
    \label{fig:all_data_qqplot_with_goog}
\end{figure}
Figure \ref{fig:all_data_qqplot_with_goog} is the QQ plot for simulations' prices(10 seconds) generated with all five groups of setup, providing additional insights to Figure \ref{fig:rl_zi_qqplot_with_goog}. More details in section 5.1.

\begin{figure}[!htb]
    \centering
    \begin{subfigure}[b]{0.4\textwidth}
        \includegraphics[width=\textwidth]{exp_img/train_autocorr_with_goog.png}
        \caption{Training Group}
        \label{fig:testing_autocorr}
    \end{subfigure}
    \begin{subfigure}[b]{0.4\textwidth}
        \includegraphics[width=\textwidth]{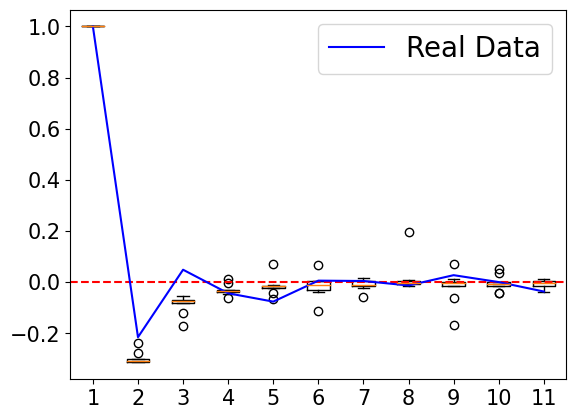}
        \caption{Testing Group}
        \label{fig:testing_autocorr}
    \end{subfigure}
    \begin{subfigure}[b]{0.4\textwidth}
        \includegraphics[width=\textwidth]{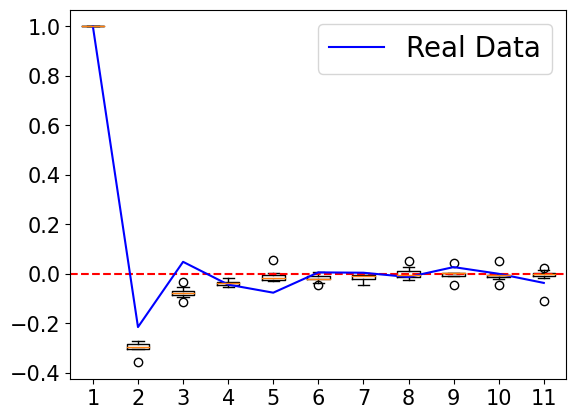}
        \caption{Untrained Group}
        \label{fig:untrained_autocorr}
    \end{subfigure}
    \begin{subfigure}[b]{0.4\textwidth}
        \includegraphics[width=\textwidth]{exp_img/ZI_autocorr_with_goog.png}
        \caption{ZI Simulation}
        \label{fig:testing_autocorr}
    \end{subfigure}
    \captionsetup{font=large}
    \caption{Auto-correlation comparison between Testing and Untrained}
    \label{fig:testing_untrained_autocorr_analysis}
\end{figure}
Figure \ref{fig:testing_untrained_autocorr_analysis} shows the ACF graphs for the price returns from groups of testing and untrained. More details in section 5.1.

\begin{figure}[!htb]
  \centering
  \begin{subfigure}[b]{0.4\textwidth}
    \centering
    \includegraphics[width=\textwidth]{exp_img/absolue_autocorr_train.png}
    \caption{Training Group}
    \label{fig:absolue_autocorr_testing}
  \end{subfigure}
  \begin{subfigure}[b]{0.4\textwidth}
    \centering
    \includegraphics[width=\textwidth]{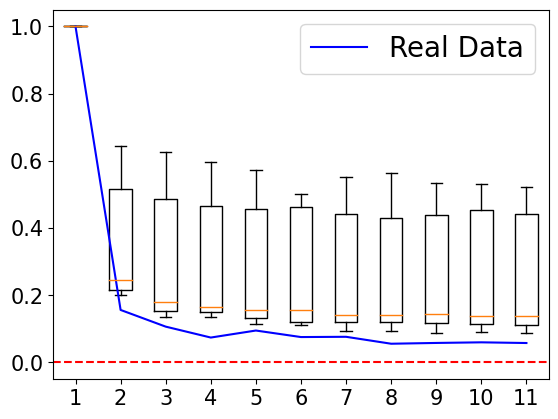}
    \caption{Testing Group}
    \label{fig:absolue_autocorr_testing}
  \end{subfigure}
  \begin{subfigure}[b]{0.4\textwidth}
    \centering
    \includegraphics[width=\textwidth]{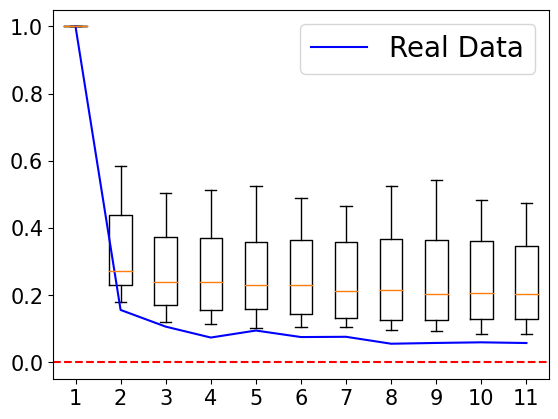}
    \caption{Untrained Group}
    \label{fig:absolue_autocorr_untrained}
  \end{subfigure}
  \begin{subfigure}[b]{0.4\textwidth}
    \centering
    \includegraphics[width=\textwidth]{exp_img/absolue_autocorr_ZI.png}
    \caption{ZI Simulation}
    \label{fig:absolue_autocorr_untrained}
  \end{subfigure}
  \captionsetup{font=large}
  \caption{Auto Correlation of Absolute Returns for Testing and Untrained}
  \label{fig:Absolute return auto correlation anaylysis for Testing and Untrained}
\end{figure}
Figure \ref{fig:Absolute return auto correlation anaylysis for Testing and Untrained} shows the ACF graphs for the absolute price returns from groups of testing and untrained. More details in section 5.1.

\begin{figure}[!htb]
    \centering

     \begin{subfigure}[b]{0.4\textwidth}
        \includegraphics[width=\textwidth]{exp_img/vol_cluster_train.png}
        \caption{Training Group}
        \label{fig:vol_cluster_testing}
    \end{subfigure}
     \begin{subfigure}[b]{0.4\textwidth}
        \includegraphics[width=\textwidth]{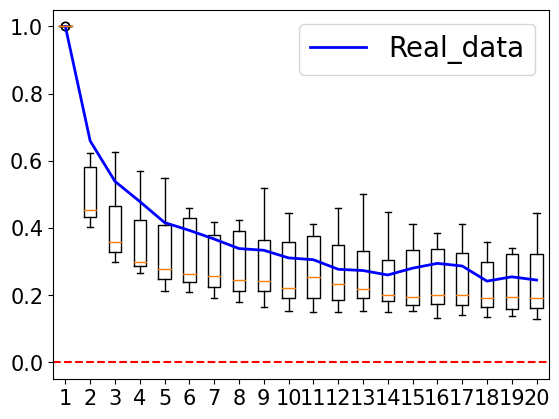}
        \caption{Testing Group}
        \label{fig:vol_cluster_testing}
    \end{subfigure}
    \begin{subfigure}[b]{0.4\textwidth}
        \includegraphics[width=\textwidth]{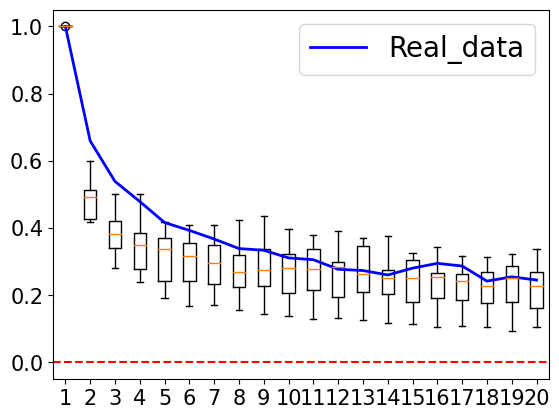}
        \caption{Untrained Group}
        \label{fig:vol_cluster_untrained}
    \end{subfigure}
    \begin{subfigure}[b]{0.4\textwidth}
        \includegraphics[width=\textwidth]{exp_img/vol_cluster_ZI.png}
        \caption{ZI Simulation}
        \label{fig:vol_cluster_untrained}
    \end{subfigure}
    \captionsetup{font=large}
    \caption{Volatility Clustering Analysis for Testing and Untrained}
    \label{fig:Volatility Clustering Analysis for Testing and Untrained}
\end{figure}
Figure \ref{fig:Volatility Clustering Analysis for Testing and Untrained} shows the volatility clustering analysis graphs for groups of testing and untrained. The analysis method can be found in \cite{cont2001empirical}. More details in section 5.1. 

\begin{table}[!htb]
  \centering
  \captionsetup{font=large}
  \caption{Returns' kurtosis and MM's inventory risk}
    \begin{tabular}{|c|c|c|c|}
      \hline
       & Continual Train & Testing & Non Train \\
      \hline
      Kurtosis & 10.24 & 142.65 & 74.25 \\
      \hline
      Inventory Risk & 3660.89 & 4565.03 & 3698.15 \\
      \hline
    \end{tabular}%
    \label{tab:Returns' kurtosis and MM's inventory risk}
\end{table}
Table \ref{tab:Returns' kurtosis and MM's inventory risk} provides additional market characteristics(Kurtosis and Inventory Risk) for groups of continual train, testing, and non train.

\subsection{Simulation Configuration}
\begin{table}[ht]
  \centering
  \captionsetup{font=large,justification=centering}
  \caption{Agents Setup for Groups Training, Testing, and Untrained \\ 
  (The definitions of the parameters can be found in Sections 3.2.1 and 3.2.2)}
      \begin{tabular}{|c|c|}
        \hline
        & \textbf{Agent configuration}\\
        \hline
         \textbf{MM Agent 1} & $\omega$ = 0.5, $\gamma$ = 0.15,  $\alpha$ = 0.09, $P_{t}^{*, i}$ = 0.5, $\epsilon_s$ range = [-1,1]\\
        \hline
        \textbf{MM Agent 2} & $\omega$ = 0.5, $\gamma$ = 0.15,  $\alpha$ = 0.09, $P_{t}^{*, i}$ = 0.5, $\epsilon_s$ range = [-1,1]\\
        \hline
        \textbf{MM Agent 3} & $\omega$ = 0.5, $\gamma$ = 0.15,  $\alpha$ = 0.09, $P_{t}^{*, i}$ = 0.5, $\epsilon_s$ range = [-1,1]\\
        \hline
        \textbf{MM Agent 4} & $\omega$ = 0.5, $\gamma$ = 0.15,  $\alpha$ = 0.09, $P_{t}^{*, i}$ = 1, $\epsilon_s$ range = [-1,2]\\
        \hline
        \textbf{LT Agent 1} & $\omega$ = 0.5, $\gamma$ = 0.9,  $\alpha$ = 0.01, $f_{\text{buy}}^{*, i}$ = 0.2, $f_{\text{sell}}^{*, i}$ = 0.8, order size = 18\\
        \hline
        \textbf{LT Agent 2} & $\omega$ = 0.5, $\gamma$ = 0.9,  $\alpha$ = 0.01, $f_{\text{buy}}^{*, i}$ = 0.2, $f_{\text{sell}}^{*, i}$ = 0.6, order size = 18\\
        \hline
        \textbf{LT Agent 3} & $\omega$ = 0.5, $\gamma$ = 0.9,  $\alpha$ = 0.01, $f_{\text{buy}}^{*, i}$ = 0.5, $f_{\text{sell}}^{*, i}$ = 0.5, order size = 18\\
        \hline
        \textbf{LT Agent 4} & $\omega$ = 0.5, $\gamma$ = 0.9,  $\alpha$ = 0.01, $f_{\text{buy}}^{*, i}$ = 0.6, $f_{\text{sell}}^{*, i}$ = 0.4, order size = 18\\
        \hline
        \textbf{LT Agent 5} & $\omega$ = 0.5, $\gamma$ = 0.9,  $\alpha$ = 0.01, $f_{\text{buy}}^{*, i}$ = 0.8, $f_{\text{sell}}^{*, i}$ = 0.2, order size = 18\\
        \hline
        \textbf{LT Agent 6} & $\omega$ = 0.5, $\gamma$ = 0.9,  $\alpha$ = 0.01, $f_{\text{buy}}^{*, i}$ = 0.2, $f_{\text{sell}}^{*, i}$ = 0.8, order size = 18\\
        \hline
        \textbf{LT Agent 7} & $\omega$ = 0.5, $\gamma$ = 0.9,  $\alpha$ = 0.01, $f_{\text{buy}}^{*, i}$ = 0.4, $f_{\text{sell}}^{*, i}$ = 0.6, order size = 18\\
        \hline
        \textbf{LT Agent 8} & $\omega$ = 0.5, $\gamma$ = 0.9,  $\alpha$ = 0.01, $f_{\text{buy}}^{*, i}$ = 0.5, $f_{\text{sell}}^{*, i}$ = 0.5, order size = 18\\
        \hline
        \textbf{LT Agent 9} & $\omega$ = 0.5, $\gamma$ = 0.9,  $\alpha$ = 0.01, $f_{\text{buy}}^{*, i}$ = 0.6, $f_{\text{sell}}^{*, i}$ = 0.4, order size = 18\\
        \hline
        \textbf{LT Agent 10} & $\omega$ = 0.5, $\gamma$ = 0.9,  $\alpha$ = 0.01, $f_{\text{buy}}^{*, i}$ = 0.8, $f_{\text{sell}}^{*, i}$ = 0.2, order size = 18\\
        \hline
        
      \end{tabular}

  \label{tab:Agents Setup for Groups A,B,C}
\end{table}
Table \ref{tab:Agents Setup for Groups A,B,C} consists of all 14 agents' configurations for groups of training, testing, and untrained. The hyper-parameters can be referenced in section 3.2.

\begin{table}[ht]
  \centering
  \captionsetup{font=large}
  \caption{Special Setup for Flash Sale and Informed LTs}
      \begin{tabular}{|c|c|}
        \hline
                 & \textbf{Setup} \\
        \hline
        \textbf{Flash Sale Agent} &  Agent sell 30000 shares/s for 5s\\
                    &  Then wait for 400s \\
                    &   Repeat the process 88 times\\
        \hline
        \textbf{Informed LTs} &   LT takes a list of buy and sell fractions:\\
                    & Buy: [0.3, 0.4, 0.4, 0.4] \\
                    & Sell: [0.4, 0.35, 0.4, 0.3] \\
                    & Each pair of fractions lasts 10000s\\
        \hline
      \end{tabular}
  \label{tab:Special Setup for Flash Sale and Informed LTs}
\end{table}
Table \ref{tab:Special Setup for Flash Sale and Informed LTs}  describes the detailed setups for the special simulations mentioned in Section 5.2 (Flash Sale and Informed LTs).

\begin{table}[ht]
  \centering
  \captionsetup{font=large, justification = centering}
  \caption{Parameter change: Market Analysis \\
            (Observe market characteristic changes as we manipulate agents' hyper parameters)}
    \begin{tabular}{|c|c|c|c|c|}
      \hline
      & Average Spread & Price std & Average Depth & Average Volume \\
      \hline
       LT PNL & $0.147 \pm 0.129$ & $21.185 \pm 21.12$ & $8.076 \pm 2.173$ & $1832.822 \pm 503.572$ \\
      \hline
       MM PNL & $0.025 \pm 0.004$ & $0.082 \pm 0.019$ & $7.545 \pm 0.718$ & $2193.544 \pm 116.252$ \\
      \hline
       LT MM PNL & $0.024 \pm 0.004$ & $0.193 \pm 0.174$ & $7.383 \pm 0.751$ & $2206.276 \pm 75.762$ \\
      \hline
       MM MarketShare & $0.020 \pm 0.008$ & $0.102 \pm 0.021$ & $6.716 \pm 1.469$ & $1804.161 \pm 551.072$ \\
      \hline
    \end{tabular}%
  \label{tab:Parameter change: Market Analysis}
\end{table}
Table \ref{tab:Parameter change: Market Analysis} shows the market characteristics of the simulations generated from different sets of hyper-parameters.

\begin{table}[ht]
  \centering
  \captionsetup{font=large}
  \caption{Parameter change Setups}
    \begin{tabular}{|c|c|}
      \hline
      & \textbf{Setup} \\
      \hline
       \textbf{LT PNL} & LT $\omega$ = 1 \\
      \hline
       \textbf{MM PNL} & MM $\omega$ = 1 \\
      \hline
       \textbf{LT MM PNL} & LT $\omega$ =  MM $\omega$ = 1 \\
      \hline
       \textbf{MM MarketShare} &  MM $\omega$ = 1\\
                        & $P_{t}^{*, i}$ = 0.5\\
      \hline
    \end{tabular}%
  \label{tab:Parameter change Setups}
\end{table}
Table \ref{tab:Parameter change Setups} shows the different setups for the simulation results.

\end{document}